\documentclass[twocolumn,aps,prb,showpacs,showkeys,amsmath,amssymb]{revtex4-1}

\usepackage{graphicx}
\usepackage{dcolumn}
\usepackage{bm}

\begin{document}

\title[Sample title]{Magneto-structural and dynamic susceptibility studies on rare-earth rich intermetallic compound}

\author{Srikanta Goswami}
\author{P. D. Babu}
\email{Corresponding Author : pdbabu@csr.res.in}

\affiliation{UGC-DAE Consortium for Scientific Research, Mumbai Centre, BARC Campus, Mumbai 400085, INDIA}
\author{R. Rawat}
\affiliation{UGC-DAE Consortium for Scientific Research, University Campus, Khandwa Road, Indore- 452001, INDIA}

\date{\today}

\begin{abstract}
Complex magneto structural behaviour of rare rich intermetallic Tb$_3$Co is reported in this study. Below the ordering temperature T$_N$ $\sim$ 84K, it undergoes a first order magnetic transition around $\sim$ 72K confirmed from specific heat and magnetization measurements. Detailed study using magnetization, specific heat, neuron diffraction and ac-susceptibility measurements suggests that the compound in the question possesses non-collinear magnetic structure. This study focusses on the temperature evolution of magnetic order at low temperatures. Neutron Diffraction clearly shows that magnetic structure remains more or less same except for decrease in moment values in the entire temperature range although  low field magnetization exhibits a transition like feature around 30K. From neutron diffraction in presence of magnetic field and zero, it was also inferred that strong spin-lattice coupling is present at lower temperature region (T $<$ 20K) compared to that for T $>$ 40K. This is the reason for the observed drop in ZFC magnetization around 30K and is not due to any change in magnetic structure. Another important but surprising result that observed is the signature of magnetic glass below 72K in frequency dependent ac-susceptibility measurements. Real part of ac-susceptibility data shows dispersion with frequency which on detailed analysis provides evidence for spin glass behaviour in the compound riding on top of non-collinear AFM order. Higher order harmonics in ac-(non-linear) susceptibilities, which are expected to show a well-defined behaviour for pure FM or AFM or spin glass or cluster glass systems, were also measured. However, the non-linear susceptibilities do not exhibit any well-known variation but show complex behaviour thus indicating the investigated compound is neither a pure AFM or FM or spin glass system but provides another evidence for non-collinear magnetic structure.

\end{abstract}
\maketitle

\section{\label{Introduction}Introduction}
Rare earth intermetallic compounds have attracted the physicists for various interesting properties, such as vide variety of magnetic phenomena, a number of magnetic transitions as a function of temperature, complex magnetic structures, field driven magneto-structural transitions, giant magnetocaloric effect (GMCE) and so on, since last five or six decades. Once such family of compounds are rare earth rich binary compounds of R$_3$T (R = rare earth, T = transition metal) type. They crystallize in Fe$_3$C type of orthorhombic structure having space group Pnma in which rare earth atom occupies 4c and 8d crystallographic sites and transition metal goes to 4c site.\cite{Strydom1970} Among these, Tb$_3$Co is one such compound that exhibits interesting properties emerging out from its crystal and magnetic structure. This compound possesses permanent magnet properties along $c$-axis with highest (within R$_3$T family) energy product up to 140 MG Oe at $\leq$ 4.2K. \cite{Baranov1998} Tb$_3$Co is found to have quite high value of magnetocaloric effect, like change in magnetic entropy is -$\Delta S_M$ = 12.5 J/Kg-K and refrigerating cooling power (RCP) is approximately 360 J/Kg for a change of magnetic field $\Delta$H = 5 Tesla.\cite{Monteiro2016} Apart from this, the compound is reported to exhibit a modulated antiferromagnetic (AFM) structure below Neel temperature (T$_N$) 82K. As temperature is decreased, the low field magnetization data of this compound displays a ferro like character from 72K to 40K and then shows a sharp drop around $\sim$ 30K to nearly very small values like AFM.  Using neutron diffraction studies, it was reported that the magnetic structure of this compound transforms to an incommensurate phase with strong ferromagnetic component along its easy axis $c$ if we go on cool the system below a critical temperature 72K.\cite{Baranov2007} and the magnetic structure was reported to be incommensurate antiferromagnetic type with wave vectors  k = (0.155,0,0) and (0.3,0.3,0) along with commensurate magnetic structure having wave vector k = (0,0,0)  for 72K $<$ T $<$  82K. Further, it was said  \cite{Baranov2007} that magnetic structure at 1.5K  to be incommensurate with k =  (0.3,0.3,0) (which is believed to be present from 65K downwards) along with commensurate phase k = (0,0,0).  However,  these studies mostly focussed on temperature regions around the phase transition at 72K to T$_N$ and at lowest temperature around 1.5K and  do not explain the sharp drop in zero-field cooled (ZFC) magnetization around 30K. Therefore, this work focuses on the magnetic state of this compound over this region and examine the nature of transition seen in magnetization around 30K. In order to achieve this objective, detailed and systematic dc magnetization, heat capacity, frequency dependent ac susceptibility and powder neutron diffraction measurements are carried out. In this work, it will be shown that the magnetic structure remains unaltered from 2K to 70K and the transition at 30K is not a true phase transition but only arises due to change in strong magneto elastic coupling at low temperature to weak coupling at higher temperatures. It will also be shown that there is magnetic glassiness riding on top of non collinear magnetic structure.

\section{\label{experimental}experimental details}
Polycrystalline samples of Tb$_3$Co compound was synthesized by arc-melting with the constituent elements taken in high purity (Tb $\sim  99.9 \% $ and Co $\sim 99.999\% $) in argon atmosphere. The ingots were re melted several times turning upside down for having better homogenisation. Phase purity of these compounds was established by X-ray diffraction and neutron diffraction. Dc magnetization measurements were carried out using commercial Quantum Design make 9-Tesla Physical Property Measurement System based vibrating sample magnetometer (VSM). Real and imaginary parts of ac susceptibility along with higher harmonics at different frequencies were measured using ACMS option of 9T PPMS. Specific heat was measured on 14 Tesla Dynacool PPMS (Quantum Design make) with heat capacity option that employs relaxation calorimetry. Neutron diffraction (ND) was performed at a wavelength of $\lambda $ = 1.48{\AA} and 2.315{\AA} using focusing crystal diffractometer (FCD-PD-III) at Dhruva reactor, BARC, India. Low temperature neutron diffraction data was collected using a close cycle refrigerator (CCR) and a 7T cryogen free magnet (Cryogenic Ltd UK make) set up.

\section{\label{results}results and discussion}
\subsection{\label{Dc Magnetization}Dc Magnetization}
The temperature dependent magnetization data for Tb$_3$Co measured in low field of 100Oe using the ‘zero-field cooled warming’ (ZFC) and ‘Field Cooled warming’ (FCW) protocols is shown in Fig.1. The ZFC curve shows three transitions; at 84K ( = T$_N$), 72K and sharp drop at $\sim$ 40K. Whereas the FCW curves coincides ZFC till 68K then deviates. There is no low temperature drop in FC curve at 40K, only slight slope change is noticed before it levels off. Above T$_N$, the ZFC data taken at 500 Oe is used to obtain inverse susceptibility ($\chi^{-1}$ = H/M) and Curie-Weiss (CW) analysis was performed based on Curie-Weiss law,

\begin{equation*}
\chi ^{-1} = \frac{T-\theta_p}{C}
\end{equation*}

\begin{figure}
\centering
\includegraphics[width=3.4in,trim=0.09in 0.30in 0.09in 0.13in]{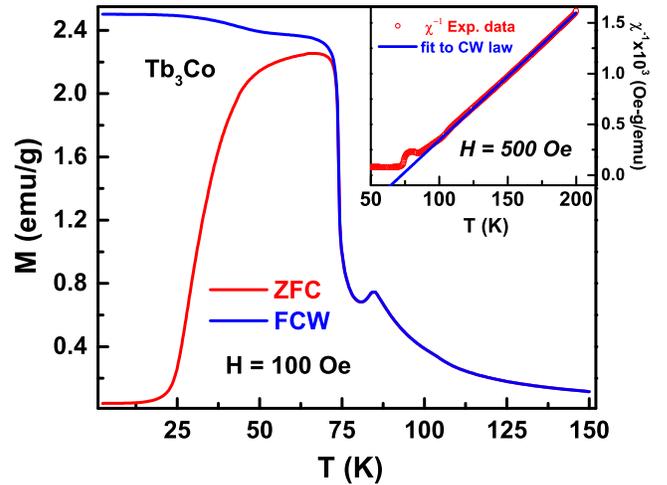}
\caption{ZFC-FCW of Tb$_3$Co. Curie-Wiess fit to paramagnetic region (inset).}\label{Fig1}
\end{figure}
\begin{figure}
  \centering
   \includegraphics[width=3.4in,trim=0.09in 0.30in 0.09in 0.13in]{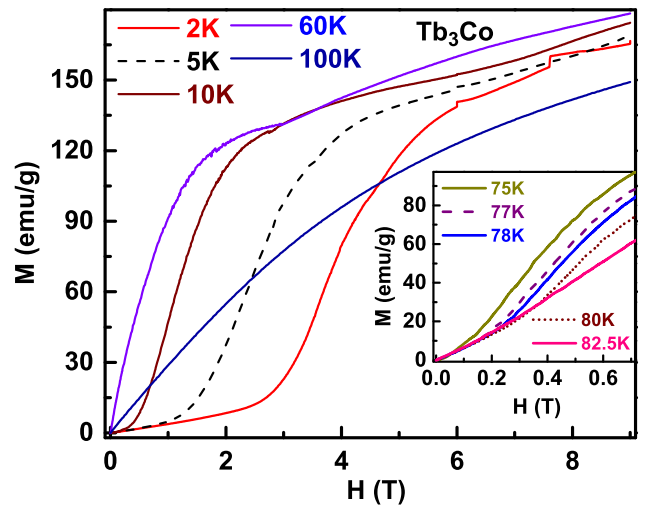}
  \caption{Virgin curves of Tb$_3$Co. Inset shows the weak signature of metamagnetic like transition in the temperature range just below T$_N$.}\label{Fig2}
\end{figure}
\begin{figure}
  \centering
   \includegraphics[width=3.4in,trim=0.09in 0.30in 0.09in 0.13in]{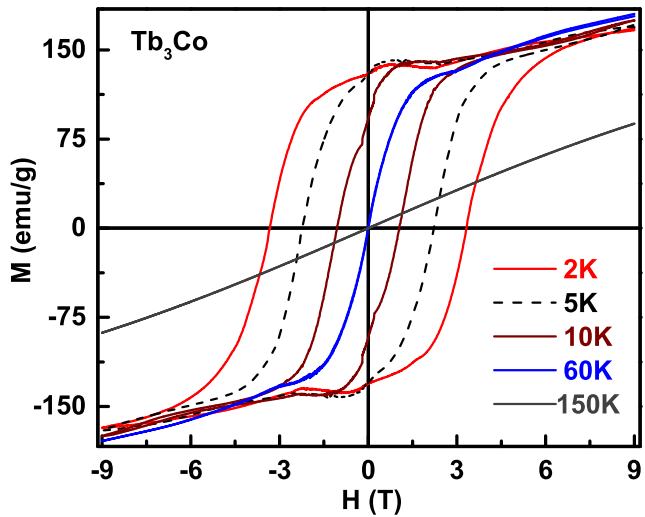}\\
  \caption{M vs H loops of Tb$_3$Co.}\label{fig3}
\end{figure}
\begin{figure}
  \centering
   \includegraphics[width=3.4in,trim=0.09in 0.30in 0.09in 0.13in]{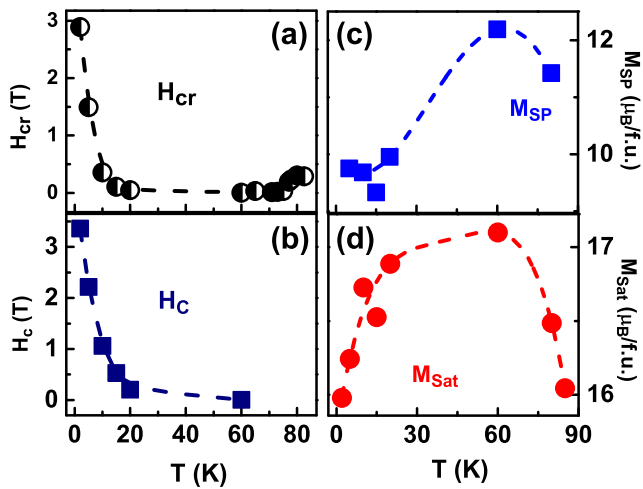}\\
  \caption{(a) elucidates the variation of critical field at which metamagnetic like transition had taken place over the temperature region. (b) shows the changes in coercivity with temperature. (c) shows the variation for spontaneous magnetization with temperature. (d) shows the same temperature variation of saturation magnetization at 9T.}\label{fig4}
\end{figure}
where C is the Curie constant and $\theta$$_P$ is the Curie-Weiss temperature.  Linear fit to CW law in the paramagnetic region as shown in the inset of Fig.1 gives $\theta$p $\sim$ 72.8K. The effective paramagnetic moment $\mu_{eff}$ obtained from Curie constant C using the relation, $\mu _{eff} = 2.828 \sqrt{CA}$, (A being the molecular weight) turns out to be  $\mu _{eff} $ $\sim$ 18.46 $\mu _B$/f.u. From this, the effective paramagnetic moment for Tb ion turns out to be $\sim$ 10.66$\mu$$_B$ assuming Co does not carry any moment in these type of compounds. \cite{Gignoux1982} However, this observed $\mu_{eff}$ is slightly higher than the free ion moment of value of 9.7$\mu$$_B$ of Tb$^3$$^+$ ion. A similar deviation of effective paramagnetic moment from free ion value was also observed for other R$_3$T compounds. \cite{Baranov2005} It was argued that deviation is due to contributions from moment coming from transition metal atoms that originate from spin fluctuations induced by the f-d exchange interactions. Figure-2 shows the virgin magnetization versus field isotherms measured up to maximum field of 9 Tesla at several selected temperatures whereas Fig.3 shows the full loops. The MH virgin curves at 2K to 20K exhibit AFM to FM metamagnetic like transition at around certain critical field H$_{cr}$, which is indicative of antiferromagnetic like order at low temperatures. This is also consistent with ZFC curve reaching nearly zero magnetization value below 20K. As the temperature is raised from 2K, the critical field   H$_{cr}$ for metamagnetic like transition, which is of the order of H$_{cr}$ $\sim$ 3T at 2K, gradually shifts to lower and lower fields and is not discernible above 20K (Fig.4 (a)). But it makes come back, though H$_{cr}$ value is small $\sim$ 0.35T, above 72K. For T $>$ 20K, no meta-magnetic like transition is observed and sample behaves like a ferromagnet. However, magnetization does not saturate in fields up to 9 Tesla. This non saturation may be because of canted spin like structure that is still prevalent even at 9T fields. According to Deryagin et al \cite{Deryagin1977} the reason behind not reaching saturation may be an extremely high value of the magnetocrystalline anisotropy energy which is comparable with the exchange interaction energy. According to them this will be reflected in high value of coercive field also. This is certainly verified by the M vs H loops which shows large coercive field at 2K (Fig.3), gradually decreasing with increasing temperature (Fig.4 (b)). Large coercive field at 2K makes this compound suitable for making a permanent magnet at this temperature, as also described by Li et al.\cite{Baranov1998, Li2008} Hereafter, the spontaneous magnetization (M$_{SP}$) obtained from linear extrapolation of high field region of MH curves below 20K turns out to be $ \sim 5.63 \mu _B$/ion and it varies in a certain manner shown in (Fig.4 (c)). The value of saturation magnetization (M$_S$$_a$$_t$) at 9T was also found to decrease with increase of temperature as shown in (Fig.4 (d)). The saturation magnetic moment obtained in the region below 20K is found to be $\sim  9.37 \mu _B$/ion, slightly lower than the free ion magnetic moment value of Tb. This reduction in magnetic moment value could be because of crystal electric field effect (CEF) present in the system as well as crystalline anisotropy.

\subsection{\label{Specific Heat}Specific Heat}
\begin{figure}
  \centering
   \includegraphics[width=3.4in,trim=0.09in 0.30in 0.09in 0.13in]{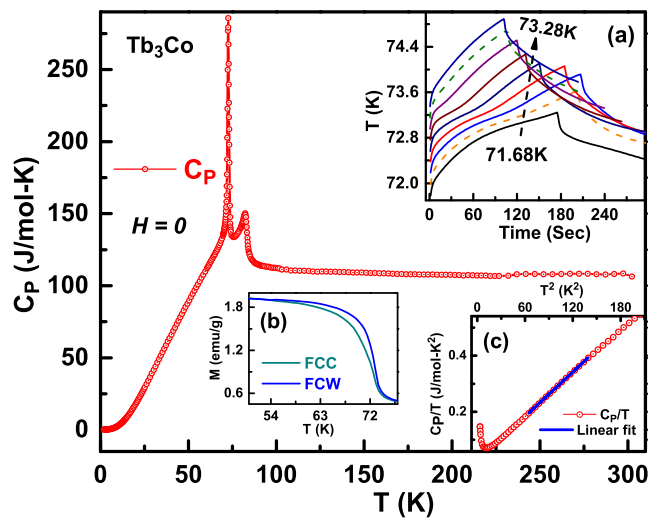}\\
  \caption{Specific heat of Tb$_3$Co. Inset (a) shows temperature vs. time plot and inset (b) displays FCC-FCW magnetization to prove a first order transition at $\sim$ 72K. Inset (c) depicts $C_P /T$ vs. T$^2$ plot along with the linear fitting.}\label{fig5}
\end{figure}
\begin{figure}
  \centering
   \includegraphics[width=3.4in,trim=0.09in 0.30in 0.09in 0.13in]{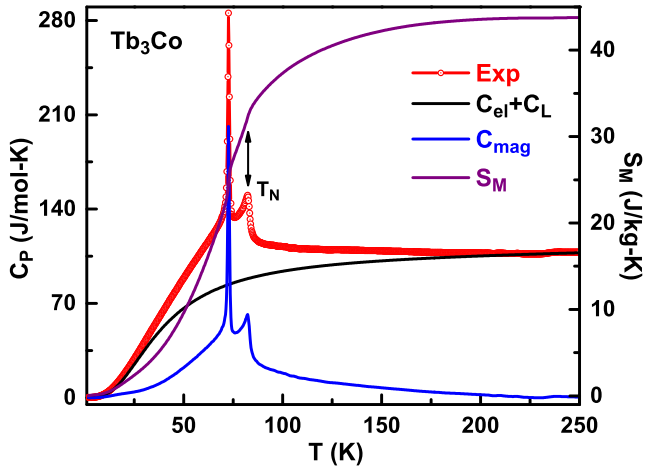}\\
  \caption{Magnetic entropy calculated for Tb$_3$Co.}\label{fig6}
\end{figure}
Figure-5 shows the zero- field specific heat (C$_P$) data as function of temperature. The C$_P$ shows a large sharp peak at 72K along with another sharp but relatively smaller peak corresponding to T$_N$ $\sim$ 84K. The transition at $\sim$ 72K is a first order phase transition of order to order (magnetic nature). Evidence for first order nature of transition was also indicated by slope change in the heat capacity relaxation curves near the transition as shown in the inset (a) of Fig.5. Slope change in relaxation curve (temperature vs. time) in two tau relaxation calorimetry is a characteristic feature of latent heat. \cite{Lashley2003} Further evidence for first order transition was provided by the thermal hysteresis ( $\sim$ 2.3K) observed between field cooled cooling (FCC) and FCW (field cooled warming) magnetization data of 100Oe as seen in the inset (b) of Fig.5.
The zero field specific heat data was analysed for electronic, lattice as well as magnetic contribution to the specific heat of the sample. The linear region in the plot of C$_P$/T vs. T$^2$ in the low temperature regime, as shown in the inset (c) of Fig.5, is analysed in terms of the equation C$_p$/T = $\gamma$T + $\beta$T$^2$. Here, $\gamma$ is Sommerfeld coefficient, related to the electronic contribution of the specific heat, C$_{el}$ = $\gamma$T, and $\beta$ is related to Debye temperature ($\Theta$$_D$). The values of $\Theta$$_D$ and $\gamma$ determined from fits turn out to be $\sim$ 143K and $\sim$ 24.3 mJ/mol-K$^2$, respectively. These values slightly differ from earlier reports \cite{Monteiro2016} because the crystal field effects at very low temperatures makes estimation of accurate Debye temperature very difficult. This was pointed out by other workers as well that accurate determination of Debye temperature for these type of compounds becomes difficult owing to low temperature anomaly and non-linearity in the C$_P$/T vs.T$^2$ data in the low temperatures region.\cite{Tristan2004}
In order to obtain magnetic entropy, Sm, from the relation $S_m$ = $\int _{0}^{T} C_{mag} / T dt$, the magnetic contribution to the specific heat C$_{mag}$ needs to be estimated by subtracting phonon and electric contributions from total specific heat. The non-magnetic specific heat contribution was calculated using full Debye expression \cite{Gopal1966} plus electronic as shown in Fig.6. The generated data (C$_{lattie}$ + C$_{el}$) as well as the experimental data matches with the Doulong petits law at higher temperature region with the value of 3nR $\sim$ 100J/mol-K. The best fit was obtained with the Debye temperature $\Theta _D$ = 158K and Sommerfeld’s coefficient remained same as obtained earlier (24.34 mJ/mol-K$^2$). By subtracting this part from total C$_P$, the magnetic contribution, C$_m$$_a$$_g$, was obtained from which magnetic entropy S$_m$ was estimated. Both these quantities are displayed in Fig.6. The magnetic entropy value at T$_N$ = 84K was found to be $\sim$ 33 J/mol-K which is consistent with that observed previously. \cite{Baranov2007,Baranov2004} The S$_m$ value at T$_N$ should be equal to S$_m$ (T = T$_N$) = 3Rln (2J+1), where R is gas constant and J is total angular momentum. This implies (2J+1) $\sim$ 4, which means only four multiplets are responsible for the magnetic behaviour in Tb$_3$Co compound as compared to all the multiplets contributing to magnetic behaviour in Tb metal (since J = 6 for Tb). This is in agreement with earlier reports. \cite{Baranov2007,Baranov2004}

\subsection{\label{Neutron Diffraction}Neutron Diffraction}

\begin{figure}
  \centering
   \includegraphics[width=3.4in,trim=0.09in 0.30in 0.09in 0.13in]{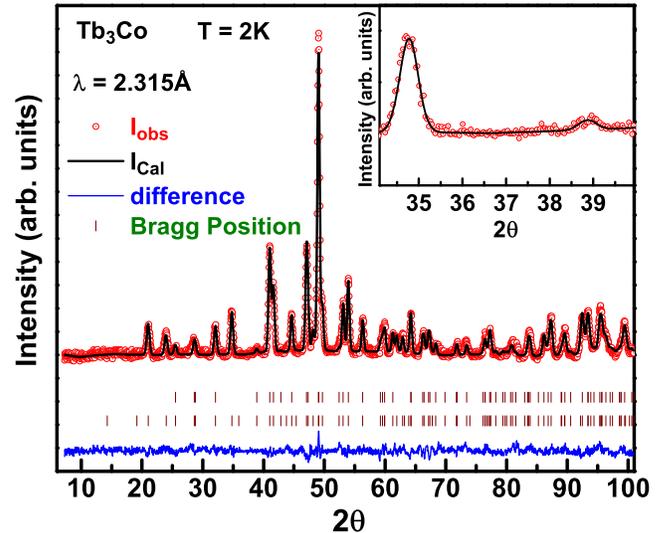}\\
  \caption{Rietveld refined neutron diffraction pattern of Tb$_3$Co at 2K. Inset shows zoomed portion of 2$\theta$ range of 34$^\circ$ to 40$^\circ$.}\label{fig7}
\end{figure}
\begin{figure}
  \centering
   \includegraphics[width=3.4in,trim=0.09in 0.30in 0.09in 0.13in]{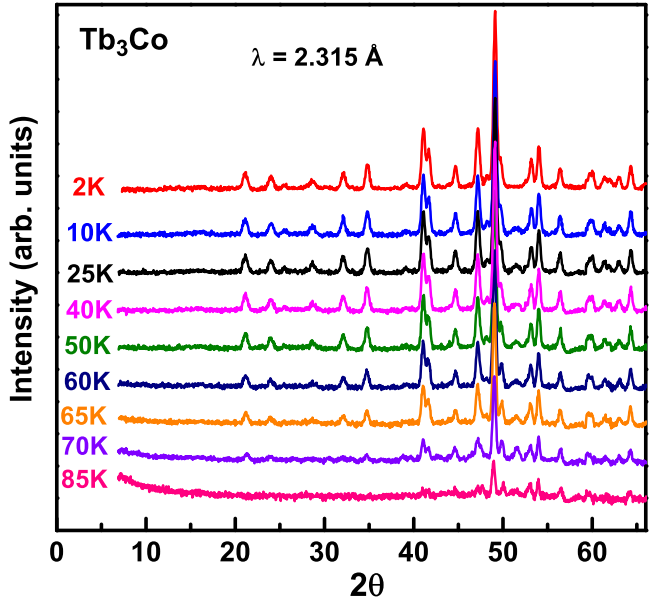}\\
  \caption{Temperature dependent variation of ND pattern of Tb$_3$Co.}\label{fig8}
\end{figure}
\begin{figure}
  \centering
   \includegraphics[width=3.4in,trim=0.09in 0.30in 0.09in 0.13in]{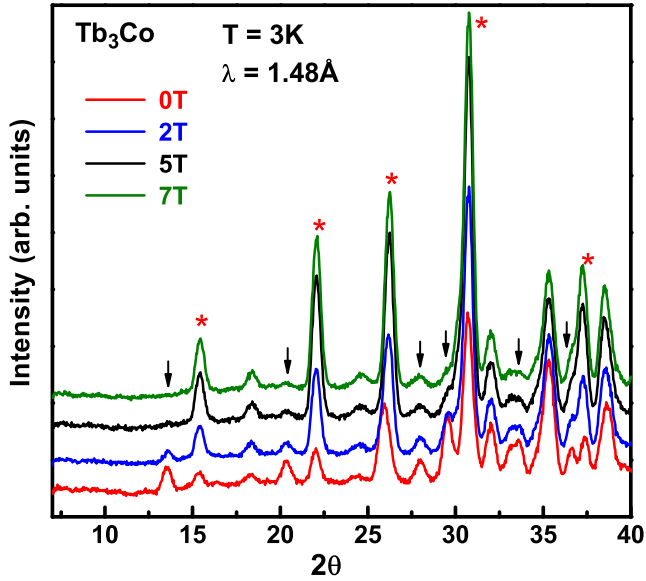}\\
  \caption{ND patterns at different magnetic fields. AFM peaks are indicated by arrows and FM peaks are denoted by star.}\label{fig9}
\end{figure}
\begin{figure}
  \centering
   \includegraphics[width=3.4in,trim=0.09in 0.30in 0.09in 0.13in]{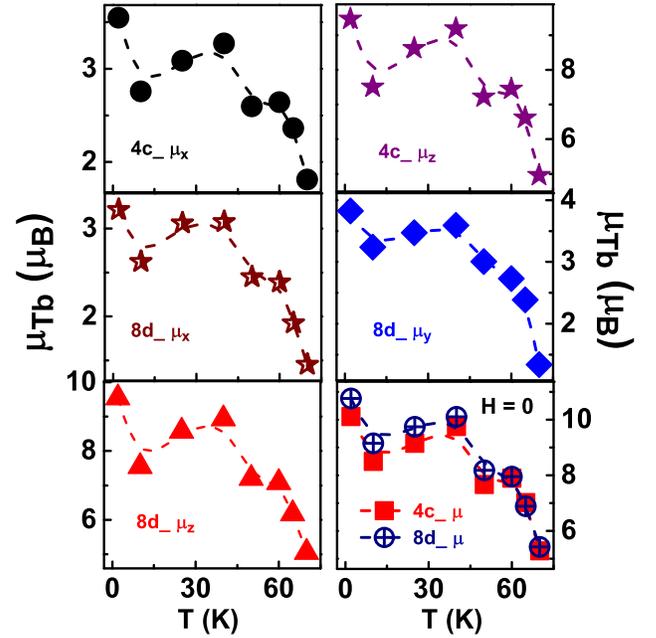}\\
  \caption{Variation of Tb moments of individual sites along with the total moments in Tb$_3$Co.}\label{fig10}
\end{figure}
\begin{figure}
  \centering
   \includegraphics[width=3.4in,trim=0.09in 0.30in 0.09in 0.13in]{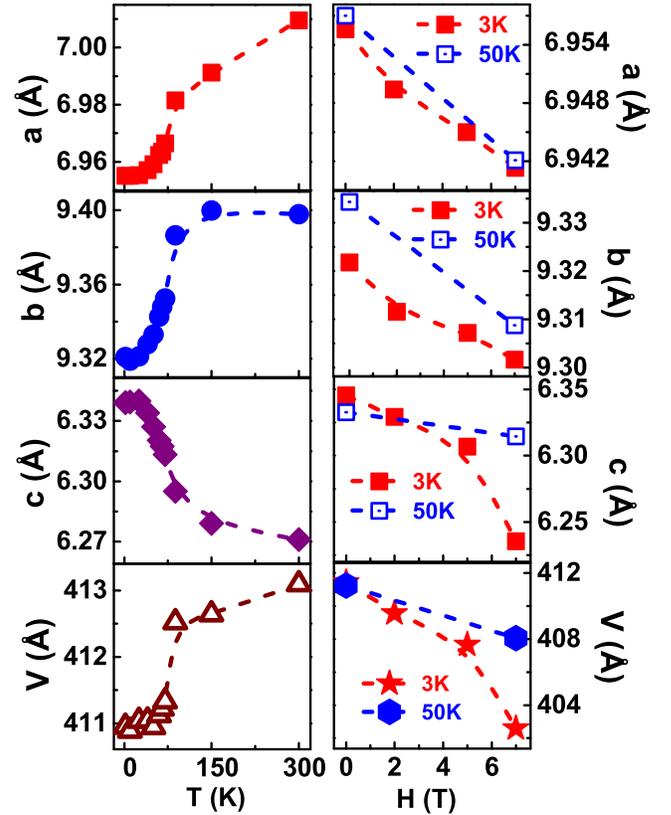}\\
  \caption{The left panel shows the variation of lattice constants and volume with temperature whereas the right panel shows the same with field.}\label{fig11}
\end{figure}
In order to understand the microscopic magnetic structure of this compound temperature dependent neutron diffraction measurements were carried out at wavelength of 1.48{\AA} and 2.315{\AA}. Rietveld refinement of room temperature ND pattern using Fullprof program \cite{Carvajal1993} confirms the crystal structure to be orthorhombic with ${Pnma}$ space group. The lattice parameters ($a$ = 7.0174{\AA}, $b$ = 9.4431{\AA}, and $c$ = 6.2876{\AA}) and other structural parameters obtained from the refinement are consistent with earlier reports.\cite{Buschow1969} Figure-7 shows the ND pattern recorded at 2K with a wavelength of $\lambda$ = 2.315{\AA} and Rietveld refined magnetic structure using Fullprof. High angle data of $\lambda$ = 1.48{\AA} diffraction pattern was used to refine low temperature cell parameters and other structural parameters that vary with temperature. Magnetic structure was then refined by fixing the structural parameters at these values. The magnetic structure could be refined using magnetic propagation wave vector k = (0,0,0) which is commensurate with crystal lattice. The magnetic moments obtained are $\mu$$_x$ = -3.548$\mu$$_B$, $\mu$$_y$ = 0 and $\mu$$_z$ = 9.471$\mu$$_B$ for Tb at 4c site and $\mu$$_x$ = 3.213$\mu$$_B$, $\mu$$_y$ = 3.825$\mu$$_B$ and $\mu$$_z$ = 9.541$\mu$$_B$ for Tb at 8d site. No discernible magnetic moment was found on the transition element, Co, in this compound. The magnetic refinement at 2K is shown in Fig.7. This result is consistent with that reported by Gignoux et al \cite{Gignoux1982} and Baranov et al.\cite{Baranov2007} However, Baranov et al. reported in addition to magnetic propagation vector k = (0,0,0) another incommensurate structure corresponding k = (0.3,0.3,0) in their magnetic structure analysis of neutron diffraction taken at a temperature of T = 1.5K. They have invoked this incommensurate structure to fit three very small peaks found in their data over the 2$\theta $ range of $\sim$ 60$^\circ $ to $\sim$ 70$^\circ $ corresponding to a  $\lambda$ = 3.88{\AA}. This angular range of 2$\theta$ for our wavelength of $\lambda$ = 2.315{\AA} would correspond to $\sim$ 34.7$^\circ $ to $\sim$ 40$^\circ $. Inset of Fig.7 shows the zoomed portion of 2K neutron data in this angular range. However, our diffraction pattern does not show these peaks. If present, we should have also seen these peaks given their intensities and considering our diffractometer’s resolution and intensities. It is probable that this incommensurate phase is discernible only at 1.5K. As temperature raised from 2K, magnetization data, ZFC curve and MH virgin curves seemed to indicate that magnetic order or structure is changing around $\sim$ 30K to $\sim$ 40K and a ferromagnetic order seem to set in over $\sim$ 40K to $\sim$ 70K. The previous neutron diffraction studies did not focus over this temperature range. Therefore, temperature evolution of neutron diffraction patterns from 2K to 85K were measured as shown in Fig. 8. A critical examination of the diffraction patterns in Fig.8 shows that the patterns remain more or less same right up to 70K thereby suggesting magnetic structure remains unaltered in this temperature range.  In order to see the diffraction pattern when the system is driven to ferromagnetic state by magnetic field, neutron diffractions were recorded at $\lambda$ = 1.48{\AA} in presence of magnetic field up to 7 Tesla as shown in Fig.9. As the magnetic field is increased, certain peaks (AFM-indicated by arrows) are decreasing and nearly disappear at 7T whereas FM peaks increase in intensity. Clearly the pattern at 7T (nearly ferromagnetic-7T is not sufficient to completely align all the spins in the direction of field as indicated by magnetization) is completely different from that at any temperature shown in Fig.8 and thus ruling out pure ferromagnetic order at any temperature. Detailed Rietveld refinements of all the neutron data above 2K up to 70K reveals that the magnetic structure broadly remains same. The magnetic moments gradually decrease with increasing temperature as shown Fig.10. The temperature variation of total 4c and 8d site moments is also shown in Fig.10. The variation of lattice parameters and volume with temperature is shown in Fig.11. The values of lattice parameter $b$ and volume does not change much up to $\sim$ 20K and then starts to increase and show saturation above 40K. The lattice constant $a$ also follows similar trend up to 40K and above 40K it exhibits a gradual increase instead of saturating. This variation almost mimics the ZFC magnetization. On the other hand, an inverse behaviour was observed for lattice constant $c$, which decreases with increase of T, drops sharply between 30K to 40K and tends to saturate above 40K. These variations of structural parameters and behaviour of magnetization indicates a presence of spin-lattice coupling in this compound. Right side panel of Fig.11 shows the variation of these structural parameters with magnetic field at 3K and 50K which were obtained from in field ND data taken at those temperatures. It is seen that with increasing field all parameters are decreasing both at 3K and 50K, the drop in values of $a$ and $b$ are only marginal ( $\sim$ 0.2\%) but the drop is more significant ( $\sim$ 1.73\%) in the case of $c$ and volume ( $\sim$ 2.14\%) at 3K. However, at 50K, the changes in $c$ and volume are very small ( $\sim$ 0.28\%, and $\sim$ 0.77\% respectively). These observations clearly indicate a strong spin-lattice coupling below 30K and a weak coupling above 40K. It is for this reason that a robust metamagnetic like behaviour was observed for T $<$ 30K where above T $>$ 40K magnetization easily orients in the direction of the magnetic field in M vs H isotherms. However, one thing that still not clear is that why did the ZFC magnetization decrease below 30-40K and nearly shows very small value below 20K when neutron diffraction is essentially showing that there no major change in magnetic structure below 20K or in the range of 40 to 70K. Is there any additional magnetic behaviour that exists and not reflected in neutron diffraction results?. In order to address this issue, ac susceptibility at different frequencies was measured on this compound. Before discussing ac-susceptibility results, it should be pointed that the magnetic structure in the temperature range from 70K to T$_N$ = 84K was well studied previously. \cite{Baranov2007} The magnetic structure, following the first order transition at 72K to T$_N$ was established to be mixture of low commensurate structure with wave vector k =(0,0,0) and incommensurate antiferromagnetic structure with k = (0.155,0,0). Therefore, no effort was made to study this temperature range (70K to T$_N$) in the present work.

\subsection{\label{Ac magnetic Susceptibility}Ac magnetic Susceptibility}
\begin{figure}
  \centering
   \includegraphics[width=3.4in,trim=0.09in 0.30in 0.09in 0.13in]{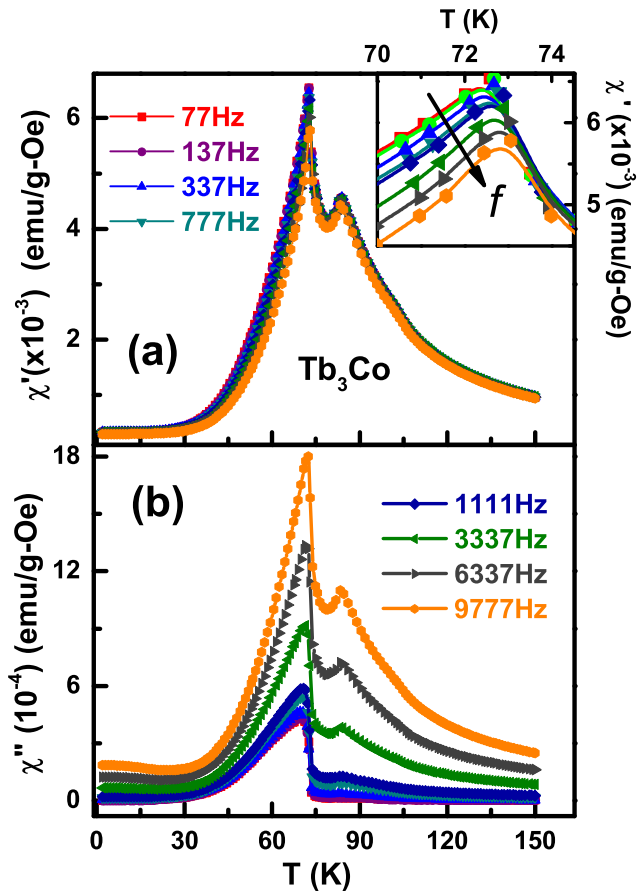}\\
  \caption{(a) The real part of ac susceptibility as a function of temperature. Inset shows the zoomed portion of the peak at T$_f$. (b) shows the temperature dependent imaginary part of ac susceptibility of Tb$_3$Co.}\label{fig12}
\end{figure}
\begin{figure}
  \centering
   \includegraphics[width=3.4in,trim=0.09in 0.30in 0.09in 0.13in]{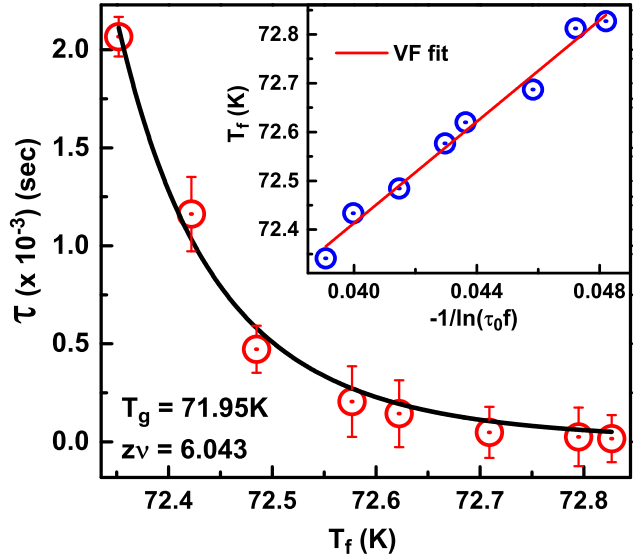}\\
  \caption{Critical slowing down of the system. Inset shows the fitting to Vogel-Fulcher law.}\label{fig13}
\end{figure}
\begin{figure}
  \centering
   \includegraphics[width=3.4in,trim=0.09in 0.30in 0.09in 0.13in]{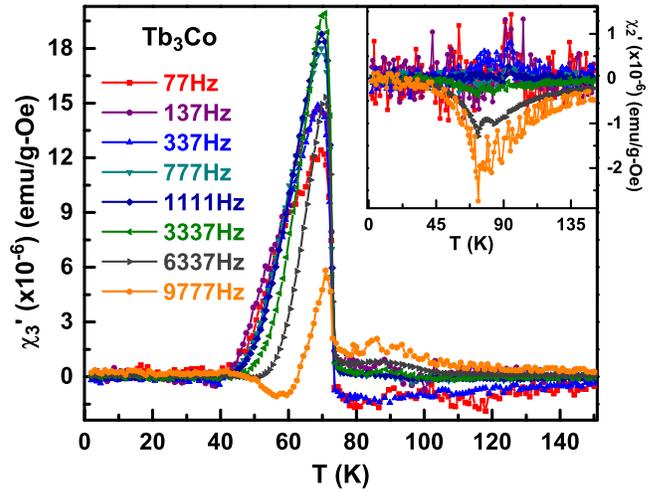}\\
  \caption{Higher harmonics $\chi _2\prime$, $\chi _3\prime$ of Tb$_3$Co as a function of temperature.}\label{fig14}
\end{figure}
\begin{figure}
  \centering
   \includegraphics[width=3.4in,trim=0.09in 0.30in 0.09in 0.13in]{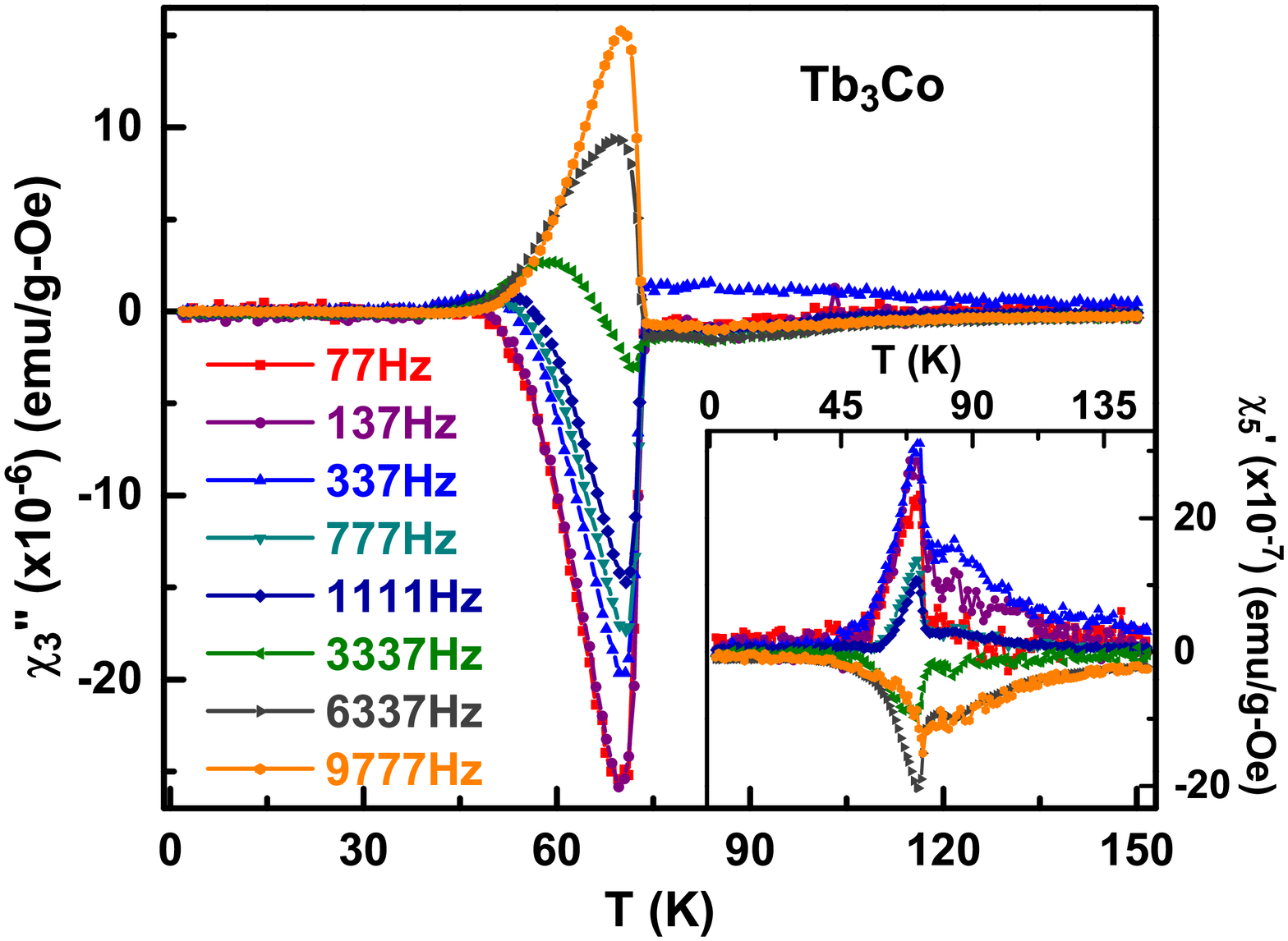}\\
  \caption{Temperature dependence of $\chi _3\prime\prime$, $\chi _5\prime$ of Tb$_3$Co.}\label{fig15}
\end{figure}
Figure-12 (a) and (b) show the real and imaginary parts of ac susceptibility on Tb$_3$Co measured with ac field amplitude of 3Oe and at frequencies ($f$) 77, 137, 337, 777, 1111, 3337, 6337, 9777 Hz.
Two peaks were seen in both $\chi \prime $ and $\chi \prime \prime $ corresponding to first order magnetic transition at 72K and T$_N$. What was surprising is that the compound exhibits dispersion in ac susceptibility below 72K. In case of spin glass systems, the first peak in $\chi \prime$ at T$_f$, is the temperature at the which spins freeze and T$_f$ shifts to higher temperature with increasing frequency as clearly seen in the inset of Fig. 12 (a) although the shift is quite small. Frequency dependency in ac-susceptibility is also exhibited by super paramagnetic clusters or particles. However, the shift in T$_f$ with $f$ is much larger compared to that of spin glasses. This can be better analysed to elucidate the nature of state below T$_f$ using the expression \cite{Sullow1997}

\begin{equation*}
\phi = \frac{\Delta T_f}{T_f \Delta (log f)}
\end{equation*}

where $\phi $ is the Mydosh parameter. The value of $\phi$ is useful in characterizing the nature of glassy behaviour. It ranges from 0.005 to 0.01 for spin glasses and 0.03 to 0.06 for cluster spin glasses. Where the value of $\phi$ is much larger for super paramagnetic systems as it assumes values 0.1 or greater. \cite{Dorman1988} In the present case the value of turns out to be 0.0032, which is slightly less than the usual values reported for spin glasses.\cite{Dorman1988,Mydosh1993} The observed glassy nature can be examined using dynamical scaling analysis called the “critical slowing down model”. In general, the spin-spin correlation length ($\xi$) diverges as $\xi$ $\sim$ $\epsilon^{-\nu}$ as T$_g$ approached from above. Here $\epsilon$ = (T - T$_g$)/T$_g$ is the reduced temperature and $\nu$ is the static critical exponent. Assuming the conventional critical slowing down on approaching T$_g$ from high temperatures, the relaxation time ($\tau$ = 2$\pi$/$f$) due to the correlated dynamics is related to $\xi$ as $\tau$ $\propto$ $\xi^z$, where z is the dynamic critical exponent. According to this model the relaxation time can be expressed \cite{Dorman1988,Hohenberg1977} as

\begin{equation*}
\tau = \tau_0 \left[ \frac{T_f - T_g}{T_g} \right]^{-z\nu }
\end{equation*}

where $\tau$ portrays the dynamical fluctuation time scale and compares to the perception time $t_{obs}$ = 1/2$\pi$$f$.  T$_f$ is the frequency dependent freezing temperature recorded from the peak value of real-part of ac susceptibility, $\chi \prime$. $\tau _0$ describes the shortest microscopic flipping time of the spins and it should be $\sim  10^{-13}$ sec close to the single spin flipping time.\cite{Gunnarsson1988,Laiho2001,Mattsson1995} The fitting of ac susceptibility data in terms of above equation is shown in Fig.13. This analysis gives $\tau _0$ =  3.7 x 10$^{-16}$ sec. which is much less than the expected value. The value of T$_g$ and z$\nu$ turn out to be T$_g$ = 72.0(10)K and z$\nu$ = 6.0(1). The value of z$\nu$ lies within the range of 5 to 10 that is expected for the different spin glass like systems. \cite{Souletie1985} The glassy spin dynamics can be further examined in terms of Neel-Arrhenius formulism for non-interacting spin systems and/or Vogel-Fulcher formulism for interacting spin systems. The frequency dependence of $T_f$ can be analysed as per Neel-Arrhenius law for thermally activated energy barriers E$_a$ in terms of the expression \cite{Pejakovic2000}, $\tau  = \tau_0 \exp{(E_a / k_B T_f)}$ where ($\tau$ = 1/f) is the relaxation time, is the characteristic relaxation time or attempt time for a single spin-flip. The fits to Neel-Arrhenius law yield unphysical values for activation energy, E$_a$ and $\tau_0$. Henceforth, an attempt was made to fit the data with Vogel-Fulcher (VF) law \cite{Gornicka2018}, $\tau  = \tau_0 \exp{ \left( E_a / k_B \left( T_f - T_0 \right) \right) }$, where $T_0$ is measure of interaction strength between the spins. The analysis was carried out by rearranging the VF equation as $T_f  = T_0 - \left( E_a / k_B \right) \left[ 1 / \ln (\tau_0 f ) \right] $, and by taking $\tau _0$ as single spin flipping time $\tau _0$ = 10$ ^{-13}$ sec. \cite{Laiho2001} This VF analysis gives a good fit (inset of Fig.13) with values for T$_0$ $\sim$ 70.3K and E$_a$/k$_B \sim$ 52.0K. The value of E$_a$/k$_B$ so obtained compares very well with another Tb rich rare earth intermetallic Tb$_5$Pd$_2$ as well as other spin glass compounds. \cite{Gubkin2013} The higher value of T$_0$ than E$_a$/k$_B$ suggests the presence of strong coupling between the spins. Further, Tholence criteria, \cite{Tholence1984} $\delta T_{Th} = \left( T_f - T_0 \right) /T_f$ was also checked for this system and it is found that for lowest frequency (77Hz corresponding T$_f$ = 72.34K) $\delta T_{T h}$ comes out to be 0.0277 whereas for highest frequency (9777Hz corresponding T$_f$ = 72.83K) the value becomes 0.0342. These values are well close to the RKKY spin glass systems,\cite{Tholence1984} thereby indicating the magnetic glassiness observed in this compound is arising to RKKY interactions present over and above direct exchange interactions. Further, non-linear susceptibilities were also measured to gain more insight into the magnetic behaviour of the investigated compound. Non-linear susceptibilities are defined in the expansion of magnetization (m) in terms of weak external magnetic field (h) expressed as\cite{Suzuki1977,Wada1980,Fujiki1981,Sato1981}

\begin{equation*}
m = m_0 + \chi _1 h + \chi _2 h^2+ \chi _3 h^3 + \chi _4 h^4+ \chi _5 h^5….
\end{equation*}

Where, $\chi _1$ is the linear susceptibility, normally denoted by $\chi$ ( = $\chi \prime $ + i$\chi \prime \prime $) without the subscript (discussed above). $\chi _2$, $\chi _3$, $\chi _4$, etc., are the non-linear susceptibilities and m$_0$ is the spontaneous magnetization. The even harmonics in the above expression, i.e., $\chi _2$, $\chi_4$, etc., would be zero because m has inversion symmetry with respect to sign change of h and m$_0$ = 0 for paramagnets and spin glasses. \cite{Bitla2012} Only odd terms are present for canonical spin glasses. However, if spontaneous magnetization or any internal field is present then the even terms ($\chi _2$, $\chi _4$,..) may be finite.  For ferromagnets, $\chi _2$ and $\chi _4$ are negative and diverge as T$_C$ is approached from T $<$ T$_C$ and suddenly becomes zero for T $\geq$ T$_C$. \cite{Bitla2011} For pure AFM order, $\chi$ shows a peak at T$_C$ and $\chi _3$ shows \cite{Fujiki1981} divergence for T $ < $ T$_N$ but above T$_N$, $\chi_3$ abrupt drops to a positive (Z $\leq $ 6) or negative value (Z $\geq $ 7) depending on the coordination number ‘Z’ and exhibits a gap at T$_N$. On the other hand for pure SG phase, odd non- linear susceptibilities will be present and both $\chi _3$ and $\chi _5$ should diverge negatively as T$_g$ approached from either side. \cite{Suzuki1977,Fujiki1981,Bitla2012,Chalupa1977} Thus, in general the linear and non-linear susceptibilities help in distinguishing SG, FM, AFM orders unambiguously. Figure-14 shows the real part of 3rd harmonic, i.e, $\chi _3 \prime$ for several frequencies from 77 Hz to 9777Hz. $\chi _3 \prime$ tends to diverge for T $<$ T$_g$ but abruptly drops a small value at T$_g$. For T $>$ T$_g$, it exhibits is slight negative values at lower frequencies (f $\leq$ 137 Hz) and positive values at higher frequencies (f $\geq$ 337 Hz). Even harmonics for e.g. $\chi _2 \prime$ shown in the inset of Fig.14 does not show any significant variation across the transition. Only at high frequencies, a weak negative divergence is seen. $\chi _3\prime\prime$ (Fig.15) shows a negative divergence for T $<$ T$_g$ and abruptly drops close to zero at Tg and above for $f$ $\leq$ 3337Hz and it turns positive side for f $\geq$ 6337Hz. Similarly, $\chi _3\prime\prime$ ($\chi _5 \prime$) also show (Fig.15) shift from negative (positive) divergence for $f$ $\leq$ 1111Hz to positive (negative) for $f$ $\geq$ 3337Hz. Clearly the non-linear susceptibilities are not exhibiting the kind of variation expected any of pure magnetic orders of FM or AFM or SG states.  This is understandable because the present compound possesses complex non-collinear AFM structure and that’s why $\chi _3 \prime$ and $\chi _5 \prime$ behaviour is somewhat closer towards AFM behaviour. At the same the linear susceptibility $\chi \prime$ and its analyses clearly shows present of glassy behaviour of spins. This glassiness is not a pure SG state at any temperature but it is riding the non-collinear AFM structure. This data also rules out any pure or predominant FM order at any temperature. It is not clear at this point of time why there is sign reversal in $\chi _3 \prime$, $\chi _3 \prime \prime$ and $\chi _5 \prime$  variations at certain frequencies. Similarly, the origin of spin glass behaviour observed is not clear. One possibility could be Co has weak moment that is giving raise glassy behaviour though transition elements in R$_3$T type compounds are believed to not carry any magnetic moment.

\section{\label{Conclusions}Conclusions}
Detailed and systematic investigations of rare earth rich intermetallic compound Tb$_3$Co was carried out using dc magnetization, specific heat, ac-susceptibility and neutron diffraction to mainly address evolution of magnetic structure over the temperature range from 2K to 72K. Magnetization indicates FM like order below 72K up to 40K and then followed by transition around 35K to AFM like state below 20K. Very clear field driven meta-magnetic transition was observed at low temperatures and it gradually weakens above 30K. Again a week meta-magnetic signature is seen above 70K to T$_N$. For T $<$ 20K, following meta-magnetic transition, the system exhibits huge hysteresis loop with coercive field of H$_C$ $\sim$ 3T at 2K, which gradually decrease as T is increased. Specific heat shows two transitions, one at 72K and another at T$_N$. The 72K peak is sharp and quite large, and heat capacity relaxation curves show slope changes clearing indicating it to be first order transition. Thermal hysteresis is also seen between FCC and FCW magnetization curves around 72K further supporting first order nature of this transition. No signature of transition was observed around 30K in heat capacity. Neutron diffraction data as function of temperature exhibits powder patterns that are similar to one another from 2K to 70K clearly indicating magnetic structure over this temperature range remains more or less same. Detailed Rietveld refinement confirms this observation and commensurate non collinear antiferromagnetic structure with magnetic wave vector k= (0,0,0). This is consistent with that reported by Gignoux et al \cite{Gignoux1982} at lowest temperature. Additional incommensurate AFM phase (with k = (0.3,0.3,0) observed by Baranov et al at 1.5K has not been observed in the present case in the entire range of 2K  to 70K. ND studies show increasing temperature results in only decrease in magnetic moments but the magnetic structure remains unaltered. Lattice parameters are nearly constant up to $\sim$ 20K and then ${a,b}$ and volume increase with T up to about 90K, and then $b$ saturates while $a$ and volume continue to increase gradually till 300K. On the other hand  parameter $c$ exhibits a sharp decrease from $\sim $  20K onwards and shows a gradual drop as T reaches 300K. This is somewhat opposite to the variation of parameter $a$. The sharp variations of lattice parameters at 20K and subsequent changes  up to 70K  nearly mimics ZFC magnetization curve, thereby indicating crucial role played by lattice constants for the observed variation of ZFC curve. In-field ND studies show that with increasing magnetic field, $a$ and $b$ decrease only marginally but significant drop is observed for $c$ and volume at 3K. However, at 50K, the changes in ${a, b, c}$ and volume are very small there by indicating a strong spin-lattice coupling below 30K and a weak coupling above 40K. This causes magnetization to orient in the field direction easily above 40K and gives rase to robust metamagnetic like behaviour for T $<$ 30K. This explains the drop in ZFC magnetization around 30K and this is not a phase transition. Frequency dependent ac- susceptibility studies reveal a surprising feature of spin glass behaviour with $\chi \prime$ showing a frequency dispersion with peak temperature shift with increasing frequency. Detailed analysis revealed evidence for classical spin glass order. However, higher harmonics or nonlinear susceptibilities, especially $\chi _3 \prime$ does not diverge negatively ( a requirement for SG order) it rather diverges positively for T $<$ Tg and abrupt drops to zero at $\sim$ T$_g$ and remains close to zero for T $>$ T$_g$. Similarly, $\chi _5 \prime$ also shows different behaviour than that expected for SG with positive peak at lower $f$ and negative peak at higher $f$. The $\chi _2 \prime$ which should diverge negatively for FM does not show any variation only show a week negative peak for higher frequencies. All these clearly show that there is underlying long range magnetic order and a magnetic glassy behaviour is riding on top of it. The origin of this glassy behaviour is not clear at present. It is possible that Co, which is believed not to possess any magnetic moment in R$_3$T compounds may have a moment and that is responsible for the additional spin glass behaviour.


\end{document}